\input harvmac
\input amssym

\def\unit{\relax{\rm 1\kern-.26em I}}
\def\nada{\relax{\rm 0\kern-.30em l}}
\def\tilde{\widetilde}

%\draftmode

%\def\Omega{\rho,\sigma,\nu  }

%% MACROS
\noblackbox
\def\IL{\relax{\rm I\kern-.18em L}}
\def\IH{\relax{\rm I\kern-.18em H}}
\def\IR{\relax{\rm I\kern-.18em R}}
\def\IC{\relax\hbox{$\inbar\kern-.3em{\rm C}$}}
\def\IZ{\relax\ifmmode\mathchoice
{\hbox{\cmss Z\kern-.4em Z}}{\hbox{\cmss Z\kern-.4em Z}}
{\lower.9pt\hbox{\cmsss Z\kern-.4em Z}} {\lower1.2pt\hbox{\cmsss
Z\kern-.4em Z}}\else{\cmss Z\kern-.4em Z}\fi}

\def\CR {{\cal R}}

\def\CL {{\cal L}}

\def\CO {{\cal O}}

%% MORE MACROS

\def\CO {{\cal O}}

\def\Tr{{\rm Tr}}

\font\manual=manfnt \def\dbend{\lower3.5pt\hbox{\manual\char127}}

\def\IZ{\relax\ifmmode\mathchoice
{\hbox{\cmss Z\kern-.4em Z}}{\hbox{\cmss Z\kern-.4em Z}}
{\lower.9pt\hbox{\cmsss Z\kern-.4em Z}} {\lower1.2pt\hbox{\cmsss
Z\kern-.4em Z}}\else{\cmss Z\kern-.4em Z}\fi}

\def\bar{\overline}

\def\rt2{\sqrt{2}}
\def\irt2{{1\over\sqrt{2}}}

\def\hat{\widehat}
%  \slashchar puts a slash through a character to represent contraction
%  with Dirac matrices. Use \not instead for negation of relations, and use
%  \hbar for hbar.
\def\slashchar#1{\setbox0=\hbox{$#1$}           % set a box for #1
   \dimen0=\wd0                                 % and get its size
   \setbox1=\hbox{/} \dimen1=\wd1               % get size of /
   \ifdim\dimen0>\dimen1                        % #1 is bigger
      \rlap{\hbox to \dimen0{\hfil/\hfil}}      % so center / in box
      #1                                        % and print #1
   \else                                        % / is bigger
      \rlap{\hbox to \dimen1{\hfil$#1$\hfil}}   % so center #1
      /                                         % and print /
   \fi}

\def\foursqr#1#2{{\vcenter{\vbox{
    \hrule height.#2pt
    \hbox{\vrule width.#2pt height#1pt \kern#1pt
    \vrule width.#2pt}
    \hrule height.#2pt
    \hrule height.#2pt
    \hbox{\vrule width.#2pt height#1pt \kern#1pt
    \vrule width.#2pt}
    \hrule height.#2pt
        \hrule height.#2pt
    \hbox{\vrule width.#2pt height#1pt \kern#1pt
    \vrule width.#2pt}
    \hrule height.#2pt
        \hrule height.#2pt
    \hbox{\vrule width.#2pt height#1pt \kern#1pt
    \vrule width.#2pt}
    \hrule height.#2pt}}}}
\def\psqr#1#2{{\vcenter{\vbox{\hrule height.#2pt
    \hbox{\vrule width.#2pt height#1pt \kern#1pt
    \vrule width.#2pt}
    \hrule height.#2pt \hrule height.#2pt
    \hbox{\vrule width.#2pt height#1pt \kern#1pt
    \vrule width.#2pt}
    \hrule height.#2pt}}}}
\def\sqr#1#2{{\vcenter{\vbox{\hrule height.#2pt
    \hbox{\vrule width.#2pt height#1pt \kern#1pt
    \vrule width.#2pt}
    \hrule height.#2pt}}}}

\def\figin{\epsfcheck\figin}\def\figins{\epsfcheck\figins}
\def\epsfcheck{\ifx\epsfbox\UnDeFiNeD
\message{(NO epsf.tex, FIGURES WILL BE IGNORED)}
\gdef\figin##1{\vskip2in}\gdef\figins##1{\hskip.5in}% blank space instead
\else\message{(FIGURES WILL BE INCLUDED)}%
\gdef\figin##1{##1}\gdef\figins##1{##1}\fi}
\def\DefWarn#1{}
\def\figinsert{\goodbreak\midinsert}
\def\ifig#1#2#3{\DefWarn#1\xdef#1{fig.~\the\figno}
\writedef{#1\leftbracket fig.\noexpand~\the\figno}%
\figinsert\figin{\centerline{#3}}\medskip\centerline{\vbox{\baselineskip12pt
\advance\hsize by -1truein\noindent\footnotefont{\bf
Fig.~\the\figno:\ } \it#2}}
\bigskip\endinsert\global\advance\figno by1}

%\PolchinskiDY
\lref\PolchinskiDY{
  J.~Polchinski,
  ``Scale And Conformal Invariance In Quantum Field Theory,''
Nucl.\ Phys.\ B {\bf 303}, 226 (1988)..
%%CITATION = UTTG-22-87%%
}

%\SundrumGV
\lref\SundrumGV{
  R.~Sundrum,
  ``SUSY Splits, But Then Returns,''
JHEP {\bf 1101}, 062 (2011).
[arXiv:0909.5430 [hep-th]].
%%CITATION = arXiv:0909.5430%%
}

%\IntriligatorIF
\lref\IntriligatorIF{
  K.~A.~Intriligator,
  ``IR free or interacting? A Proposed diagnostic,''
Nucl.\ Phys.\ B {\bf 730}, 239 (2005).
[hep-th/0509085].
%%CITATION = hep-th/0509085%%
}

%\HookFP
\lref\HookFP{
  A.~Hook,
  ``A Test for emergent dynamics,''
JHEP {\bf 1207}, 040 (2012).
[arXiv:1204.4466 [hep-th]].
%%CITATION = arXiv:1204.4466%%
}

%\SeibergPQ
\lref\SeibergPQ{
  N.~Seiberg,
  ``Electric - magnetic duality in supersymmetric nonAbelian gauge theories,''
Nucl.\ Phys.\ B {\bf 435}, 129 (1995).
[hep-th/9411149].
%%CITATION = hep-th/9411149%%
}

%\IntriligatorJJ
\lref\IntriligatorJJ{
  K.~A.~Intriligator, B.~Wecht,
  ``The Exact superconformal R symmetry maximizes a,''
Nucl.\ Phys.\  {\bf B667}, 183-200 (2003).
[hep-th/0304128].
%%CITATION = hep-th/0304128%%
}

%\ZamolodchikovGT
\lref\ZamolodchikovGT{
  A.~B.~Zamolodchikov,
  ``Irreversibility of the Flux of the Renormalization Group in a 2D Field Theory,''
JETP Lett.\  {\bf 43}, 730 (1986), [Pisma Zh.\ Eksp.\ Teor.\ Fiz.\  {\bf 43}, 565 (1986)].
}

%\KutasovUX
\lref\KutasovUX{
  D.~Kutasov,
  ``New results on the 'a theorem' in four-dimensional supersymmetric field theory,''
[hep-th/0312098].
%%CITATION = hep-th/0312098%%
}

%\AnselmiYS
\lref\AnselmiYS{
  D.~Anselmi, J.~Erlich, D.~Z.~Freedman, A.~A.~Johansen,
  ``Positivity constraints on anomalies in supersymmetric gauge theories,''
Phys.\ Rev.\  {\bf D57}, 7570-7588 (1998).
[hep-th/9711035].
%%CITATION = hep-th/9711035%%
}

%\FerraraPZ
\lref\FerraraPZ{
  S.~Ferrara, B.~Zumino,
  ``Transformation Properties of the Supercurrent,''
Nucl.\ Phys.\  {\bf B87}, 207 (1975).
%%CITATION = CERN-TH-1947%%
}

%\KomargodskiRB
\lref\KomargodskiRB{
  Z.~Komargodski, N.~Seiberg,
  ``Comments on Supercurrent Multiplets, Supersymmetric Field Theories and Supergravity,''
JHEP {\bf 1007}, 017 (2010).
[arXiv:1002.2228 [hep-th]].
%%CITATION = arXiv:1002.2228%%
}

%\AppelquistHR
\lref\AppelquistHR{
  T.~Appelquist, A.~G.~Cohen, M.~Schmaltz,
  ``A New constraint on strongly coupled gauge theories,''
Phys.\ Rev.\  {\bf D60}, 045003 (1999).
[arXiv:hep-th/9901109 [hep-th]].
%%CITATION = SLAC-PUB-8045%%
}

%\AbelWV
\lref\AbelWV{
  S.~Abel, M.~Buican and Z.~Komargodski,
  ``Mapping Anomalous Currents in Supersymmetric Dualities,''
Phys.\ Rev.\ D {\bf 84}, 045005 (2011).
[arXiv:1105.2885 [hep-th]].
%%CITATION = arXiv:1105.2885%%
}

%\KutasovIY
\lref\KutasovIY{
  D.~Kutasov, A.~Parnachev, D.~A.~Sahakyan,
  ``Central charges and U(1)(R) symmetries in N=1 superYang-Mills,''
JHEP {\bf 0311}, 013 (2003).
[hep-th/0308071].
%%CITATION = hep-th/0308071%%
}

%\PolandWG
\lref\PolandWG{
  D.~Poland, D.~Simmons-Duffin,
  ``Bounds on 4D Conformal and Superconformal Field Theories,''
JHEP {\bf 1105}, 017 (2011).
[arXiv:1009.2087 [hep-th]].
%%CITATION = arXiv:1009.2087%%
}

%\KomargodskiVJ
\lref\KomargodskiVJ{
  Z.~Komargodski and A.~Schwimmer,
  ``On Renormalization Group Flows in Four Dimensions,''
JHEP {\bf 1112}, 099 (2011).
[arXiv:1107.3987 [hep-th]].
%%CITATION = arXiv:1107.3987%%
}

%\AntoniadisGN
\lref\AntoniadisGN{
  I.~Antoniadis and M.~Buican,
  ``On R-symmetric Fixed Points and Superconformality,''
Phys.\ Rev.\ D {\bf 83}, 105011 (2011).
[arXiv:1102.2294 [hep-th]].
%%CITATION = arXiv:1102.2294%%
}

%\LutyWW
\lref\LutyWW{
  M.~A.~Luty, J.~Polchinski and R.~Rattazzi,
  ``The $a$-theorem and the Asymptotics of 4D Quantum Field Theory,''
[arXiv:1204.5221 [hep-th]].
%%CITATION = arXiv:1204.5221%%
}

%\DorigoniRA
\lref\DorigoniRA{
  D.~Dorigoni and V.~S.~Rychkov,
  ``Scale Invariance + Unitarity $=>$ Conformal Invariance?,''
[arXiv:0910.1087 [hep-th]].
%%CITATION = arXiv:0910.1087%%
}

%\NakayamaTK
\lref\NakayamaTK{
  Y.~Nakayama,
  ``Comments on scale invariant but non-conformal supersymmetric field theories,''
[arXiv:1109.5883 [hep-th]].
%%CITATION = arXiv:1109.5883%%
}

%\FortinIC
\lref\FortinIC{
  J.~-F.~Fortin, B.~Grinstein and A.~Stergiou,
  ``Scale without Conformal Invariance at Three Loops,''
[arXiv:1202.4757 [hep-th]].
%%CITATION = arXiv:1202.4757%%
}

%\FortinKS
\lref\FortinKS{
  J.~-F.~Fortin, B.~Grinstein and A.~Stergiou,
  ``Scale without Conformal Invariance: An Example,''
Phys.\ Lett.\ B {\bf 704}, 74 (2011).
[arXiv:1106.2540 [hep-th]].
%%CITATION = UCSD-PTH-11-11%%
}

%\KomargodskiXV
\lref\KomargodskiXV{
  Z.~Komargodski,
  ``The Constraints of Conformal Symmetry on RG Flows,''
[arXiv:1112.4538 [hep-th]].
%%CITATION = arXiv:1112.4538%%
}

%\MyersTJ
\lref\MyersTJ{
  R.~C.~Myers and A.~Sinha,
  ``Holographic c-theorems in arbitrary dimensions,''
JHEP {\bf 1101}, 125 (2011).
[arXiv:1011.5819 [hep-th]].
%%CITATION = arXiv:1011.5819%%
}

%\CasiniKV
\lref\CasiniKV{
  H.~Casini, M.~Huerta and R.~C.~Myers,
  ``Towards a derivation of holographic entanglement entropy,''
JHEP {\bf 1105}, 036 (2011).
[arXiv:1102.0440 [hep-th]].
%%CITATION = arXiv:1102.0440%%
}

%\JafferisZI
\lref\JafferisZI{
  D.~L.~Jafferis, I.~R.~Klebanov, S.~S.~Pufu and B.~R.~Safdi,
  ``Towards the F-Theorem: N=2 Field Theories on the Three-Sphere,''
JHEP {\bf 1106}, 102 (2011).
[arXiv:1103.1181 [hep-th]].
%%CITATION = arXiv:1103.1181%%
}

%\MyersED
\lref\MyersED{
  R.~C.~Myers and A.~Singh,
  ``Comments on Holographic Entanglement Entropy and RG Flows,''
JHEP {\bf 1204}, 122 (2012).
[arXiv:1202.2068 [hep-th]].
%%CITATION = arXiv:1202.2068%%
}

%\LiuEE
\lref\LiuEE{
  H.~Liu and M.~Mezei,
  ``A Refinement of entanglement entropy and the number of degrees of freedom,''
[arXiv:1202.2070 [hep-th]].
%%CITATION = arXiv:1202.2070%%
}

%\ElvangST
\lref\ElvangST{
  H.~Elvang, D.~Z.~Freedman, L.~-Y.~Hung, M.~Kiermaier, R.~C.~Myers and S.~Theisen,
  ``On renormalization group flows and the a-theorem in 6d,''
[arXiv:1205.3994 [hep-th]].
%%CITATION = arXiv:1205.3994%%
}

%\MaxfieldAW
\lref\MaxfieldAW{
  T.~Maxfield and S.~Sethi,
  ``The Conformal Anomaly of M5-Branes,''
[arXiv:1204.2002 [hep-th]].
%%CITATION = arXiv:1204.2002%%
}

%\JafferisUN
\lref\JafferisUN{
  D.~L.~Jafferis,
  ``The Exact Superconformal R-Symmetry Extremizes Z,''
[arXiv:1012.3210 [hep-th]].
%%CITATION = arXiv:1012.3210%%
}

%\CraigDI
\lref\CraigDI{
  N.~Craig, M.~McCullough and J.~Thaler,
  ``Flavor Mediation Delivers Natural SUSY,''
[arXiv:1203.1622 [hep-ph]].
%%CITATION = arXiv:1203.1622%%
}

%\ClossetVG
\lref\ClossetVG{
  C.~Closset, T.~T.~Dumitrescu, G.~Festuccia, Z.~Komargodski and N.~Seiberg,
  ``Contact Terms, Unitarity, and F-Maximization in Three-Dimensional Superconformal Theories,''
[arXiv:1205.4142 [hep-th]].
%%CITATION = arXiv:1205.4142%%
}

%\BuicanTY
\lref\BuicanTY{
  M.~Buican,
  ``A Conjectured Bound on Accidental Symmetries,''
Phys.\ Rev.\ D {\bf 85}, 025020 (2012).
[arXiv:1109.3279 [hep-th]].
%%CITATION = arXiv:1109.3279%%
}

%\CasiniEI
\lref\CasiniEI{
  H.~Casini and M.~Huerta,
  ``On the RG running of the entanglement entropy of a circle,''
[arXiv:1202.5650 [hep-th]].
%%CITATION = arXiv:1202.5650%%
}

%\AbelBJ
\lref\AbelBJ{
  S.~Abel and V.~V.~Khoze,
  ``Dual unified SU(5),''
JHEP {\bf 1001}, 006 (2010).
[arXiv:0909.4105 [hep-ph]].
%%CITATION = arXiv:0909.4105%%
}

%\ATLASAE
\lref\ATLASAE{
  G.~Aad {\it et al.}  [ATLAS Collaboration],
  ``Combined search for the Standard Model Higgs boson using up to 4.9 fb-1 of pp collision data at sqrt(s) = 7 TeV with the ATLAS detector at the LHC,''
Phys.\ Lett.\ B {\bf 710}, 49 (2012).
[arXiv:1202.1408 [hep-ex]].
%%CITATION = arXiv:1202.1408%%
}

%\ChatrchyanTX
\lref\ChatrchyanTX{
  S.~Chatrchyan {\it et al.}  [CMS Collaboration],
  ``Combined results of searches for the standard model Higgs boson in pp collisions at sqrt(s) = 7 TeV,''
Phys.\ Lett.\ B {\bf 710}, 26 (2012).
[arXiv:1202.1488 [hep-ex]].
%%CITATION = arXiv:1202.1488%%
}

%\DimopoulosZB
\lref\DimopoulosZB{
  S.~Dimopoulos and H.~Georgi,
  ``Softly Broken Supersymmetry and SU(5),''
Nucl.\ Phys.\ B {\bf 193}, 150 (1981)..
%%CITATION = HUTP-81/A022%%
}

%\NelsonNF
\lref\NelsonNF{
  A.~E.~Nelson and N.~Seiberg,
  ``R symmetry breaking versus supersymmetry breaking,''
Nucl.\ Phys.\ B {\bf 416}, 46 (1994).
[hep-ph/9309299].
%%CITATION = hep-ph/9309299%%
}

%\IntriligatorPY
\lref\IntriligatorPY{
  K.~A.~Intriligator, N.~Seiberg and D.~Shih,
  ``Supersymmetry breaking, R-symmetry breaking and metastable vacua,''
JHEP {\bf 0707}, 017 (2007).
[hep-th/0703281].
%%CITATION = hep-th/0703281%%
}

%\IntriligatorDD
\lref\IntriligatorDD{
  K.~A.~Intriligator, N.~Seiberg and D.~Shih,
  ``Dynamical SUSY breaking in meta-stable vacua,''
JHEP {\bf 0604}, 021 (2006).
[hep-th/0602239].
%%CITATION = hep-th/0602239%%
}

%\IntriligatorRX
\lref\IntriligatorRX{
  K.~A.~Intriligator, N.~Seiberg and S.~H.~Shenker,
  ``Proposal for a simple model of dynamical SUSY breaking,''
Phys.\ Lett.\ B {\bf 342}, 152 (1995).
[hep-ph/9410203].
%%CITATION = hep-ph/9410203%%
}

%\DimopoulosMI
\lref\DimopoulosMI{
  S.~Dimopoulos and G.~F.~Giudice,
  ``Naturalness constraints in supersymmetric theories with nonuniversal soft terms,''
Phys.\ Lett.\ B {\bf 357}, 573 (1995).
[hep-ph/9507282].
%%CITATION = hep-ph/9507282%%
}

%\DineNP
\lref\DineNP{
  M.~Dine, R.~G.~Leigh and A.~Kagan,
  ``Flavor symmetries and the problem of squark degeneracy,''
Phys.\ Rev.\ D {\bf 48}, 4269 (1993).
[hep-ph/9304299].
%%CITATION = hep-ph/9304299%%
}

%\BarbieriPD
\lref\BarbieriPD{
  R.~Barbieri, E.~Bertuzzo, M.~Farina, P.~Lodone and D.~Pappadopulo,
  ``A Non Standard Supersymmetric Spectrum,''
JHEP {\bf 1008}, 024 (2010).
[arXiv:1004.2256 [hep-ph]].
%%CITATION = arXiv:1004.2256%%
}

%\BarbieriAR
\lref\BarbieriAR{
  R.~Barbieri, E.~Bertuzzo, M.~Farina, P.~Lodone and D.~Zhuridov,
  ``Minimal Flavour Violation with hierarchical squark masses,''
JHEP {\bf 1012}, 070 (2010), [Erratum-ibid.\  {\bf 1102}, 044 (2011)].
[arXiv:1011.0730 [hep-ph]].
%%CITATION = arXiv:1011.0730%%
}

%\CraigYK
\lref\CraigYK{
  N.~Craig, D.~Green and A.~Katz,
  ``(De)Constructing a Natural and Flavorful Supersymmetric Standard Model,''
JHEP {\bf 1107}, 045 (2011).
[arXiv:1103.3708 [hep-ph]].
%%CITATION = arXiv:1103.3708%%
}

%\GherghettaWC
\lref\GherghettaWC{
  T.~Gherghetta, B.~von Harling and N.~Setzer,
  ``A Natural little hierarchy for RS from accidental SUSY,''
JHEP {\bf 1107}, 011 (2011).
[arXiv:1104.3171 [hep-ph]].
%%CITATION = arXiv:1104.3171%%
}

%\JeongEN
\lref\JeongEN{
  K.~S.~Jeong, J.~E.~Kim and M.~-S.~Seo,
  ``Gauge mediation to effective SUSY through U(1)s with a dynamical SUSY breaking, and string compactification,''
Phys.\ Rev.\ D {\bf 84}, 075008 (2011).
[arXiv:1107.5613 [hep-ph]].
%%CITATION = arXiv:1107.5613%%
}

%\EssigQG
\lref\EssigQG{
  R.~Essig, E.~Izaguirre, J.~Kaplan and J.~G.~Wacker,
  ``Heavy Flavor Simplified Models at the LHC,''
JHEP {\bf 1201}, 074 (2012).
[arXiv:1110.6443 [hep-ph]].
%%CITATION = arXiv:1110.6443%%
}

%\KatsQH
\lref\KatsQH{
  Y.~Kats, P.~Meade, M.~Reece and D.~Shih,
  ``The Status of GMSB After 1/fb at the LHC,''
JHEP {\bf 1202}, 115 (2012).
[arXiv:1110.6444 [hep-ph]].
%%CITATION = arXiv:1110.6444%%
}

%\PapucciWY
\lref\PapucciWY{
  M.~Papucci, J.~T.~Ruderman and A.~Weiler,
  ``Natural SUSY Endures,''
[arXiv:1110.6926 [hep-ph]].
%%CITATION = arXiv:1110.6926%%
}

%\BrustTB
\lref\BrustTB{
  C.~Brust, A.~Katz, S.~Lawrence and R.~Sundrum,
  ``SUSY, the Third Generation and the LHC,''
JHEP {\bf 1203}, 103 (2012).
[arXiv:1110.6670 [hep-ph]].
%%CITATION = arXiv:1110.6670%%
}

%\DelgadoKR
\lref\DelgadoKR{
  A.~Delgado and M.~Quiros,
  ``The Least Supersymmetric Standard Model,''
Phys.\ Rev.\ D {\bf 85}, 015001 (2012).
[arXiv:1111.0528 [hep-ph]].
%%CITATION = arXiv:1111.0528%%
}

%\DesaiTH
\lref\DesaiTH{
  N.~Desai and B.~Mukhopadhyaya,
  ``Constraints on supersymmetry with light third family from LHC data,''
JHEP {\bf 1205}, 057 (2012).
[arXiv:1111.2830 [hep-ph]].
%%CITATION = arXiv:1111.2830%%
}

%\NakayamaWX
\lref\NakayamaWX{
  Y.~Nakayama,
  ``Higher derivative corrections in holographic Zamolodchikov-Polchinski theorem,''
Eur.\ Phys.\ J.\ C {\bf 72}, 1870 (2012).
[arXiv:1009.0491 [hep-th]].
%%CITATION = arXiv:1009.0491%%
}

%\AkulaJX
\lref\AkulaJX{
  S.~Akula, M.~Liu, P.~Nath and G.~Peim,
  ``Naturalness, Supersymmetry and Implications for LHC and Dark Matter,''
Phys.\ Lett.\ B {\bf 709}, 192 (2012).
[arXiv:1111.4589 [hep-ph]].
%%CITATION = arXiv:1111.4589%%
}

%\AjaibHS
\lref\AjaibHS{
  M.~A.~Ajaib, T.~Li and Q.~Shafi,
  ``Stop-Neutralino Coannihilation in the Light of LHC,''
Phys.\ Rev.\ D {\bf 85}, 055021 (2012).
[arXiv:1111.4467 [hep-ph]].
%%CITATION = arXiv:1111.4467%%
}

%\IshiwataAB
\lref\IshiwataAB{
  K.~Ishiwata, N.~Nagata and N.~Yokozaki,
  ``Natural Supersymmetry and b $->$ s gamma constraints,''
Phys.\ Lett.\ B {\bf 710}, 145 (2012).
[arXiv:1112.1944 [hep-ph]].
%%CITATION = arXiv:1112.1944%%
}

%\LodoneAA
\lref\LodoneAA{
  P.~Lodone,
  ``A Motivated Non-Standard Supersymmetric Spectrum,''
[arXiv:1112.2178 [hep-ph]].
%%CITATION = arXiv:1112.2178%%
}

%\HeTP
\lref\HeTP{
  B.~He, T.~Li and Q.~Shafi,
  ``Impact of LHC Searches on Light Top Squark,''
[arXiv:1112.4461 [hep-ph]].
%%CITATION = arXiv:1112.4461%%
}

%\ArvanitakiCK
\lref\ArvanitakiCK{
  A.~Arvanitaki and G.~Villadoro,
  ``A Non Standard Model Higgs at the LHC as a Sign of Naturalness,''
JHEP {\bf 1202}, 144 (2012).
[arXiv:1112.4835 [hep-ph]].
%%CITATION = arXiv:1112.4835%%
}

%\AuzziEU
\lref\AuzziEU{
  R.~Auzzi, A.~Giveon and S.~B.~Gudnason,
  ``Flavor of quiver-like realizations of effective supersymmetry,''
JHEP {\bf 1202}, 069 (2012).
[arXiv:1112.6261 [hep-ph]].
%%CITATION = arXiv:1112.6261%%
}

%\CsakiFH
\lref\CsakiFH{
  C.~Csaki, L.~Randall and J.~Terning,
  ``Light Stops from Seiberg Duality,''
[arXiv:1201.1293 [hep-ph]].
%%CITATION = arXiv:1201.1293%%
}

%\CraigYD
\lref\CraigYD{
  N.~Craig, M.~McCullough and J.~Thaler,
  ``The New Flavor of Higgsed Gauge Mediation,''
JHEP {\bf 1203}, 049 (2012).
[arXiv:1201.2179 [hep-ph]].
%%CITATION = arXiv:1201.2179%%
}

%\LarsenRQ
\lref\LarsenRQ{
  G.~Larsen, Y.~Nomura and H.~L.~L.~Roberts,
  ``Supersymmetry with Light Stops,''
[arXiv:1202.6339 [hep-ph]].
%%CITATION = arXiv:1202.6339%%
}

%\ATLASAE
\lref\ATLASAE{
  G.~Aad {\it et al.}  [ATLAS Collaboration],
  %``Combined search for the Standard Model Higgs boson using up to 4.9 fb$^{-1}$ of $pp$ collision data at $\sqrt{s}=7$ TeV with the ATLAS detector at the LHC,''
Phys.\ Lett.\ B {\bf 710}, 49 (2012).
[arXiv:1202.1408 [hep-ex]].
%%CITATION = arXiv:1202.1408%%
}

%\CraigHC
\lref\CraigHC{
  N.~Craig, S.~Dimopoulos and T.~Gherghetta,
  ``Split families unified,''
JHEP {\bf 1204}, 116 (2012).
[arXiv:1203.0572 [hep-ph]].
%%CITATION = arXiv:1203.0572%%
}

%\AlvesFT
\lref\AlvesFT{
  D.~S.~M.~Alves, M.~R.~Buckley, P.~J.~Fox, J.~D.~Lykken and C.~-T.~Yu,
  ``Stops and MET: the shape of things to come,''
[arXiv:1205.5805 [hep-ph]].
%%CITATION = arXiv:1205.5805%%
}

%\HanFW
\lref\HanFW{
  Z.~Han, A.~Katz, D.~Krohn and M.~Reece,
  ``(Light) Stop Signs,''
[arXiv:1205.5808 [hep-ph]].
%%CITATION = arXiv:1205.5808%%
}

%\KaplanGD
\lref\KaplanGD{
  D.~E.~Kaplan, K.~Rehermann and D.~Stolarski,
  ``Searching for Direct Stop Production in Hadronic Top Data at the LHC,''
[arXiv:1205.5816 [hep-ph]].
%%CITATION = arXiv:1205.5816%%
}

%\ArkaniHamedWC
\lref\ArkaniHamedWC{
  N.~Arkani-Hamed and R.~Rattazzi,
  ``Exact results for nonholomorphic masses in softly broken supersymmetric gauge theories,''
Phys.\ Lett.\ B {\bf 454}, 290 (1999).
[hep-th/9804068].
%%CITATION = hep-th/9804068%%
}

%\GreenDA
\lref\GreenDA{
  D.~Green, Z.~Komargodski, N.~Seiberg, Y.~Tachikawa and B.~Wecht,
  ``Exactly Marginal Deformations and Global Symmetries,''
JHEP {\bf 1006}, 106 (2010).
[arXiv:1005.3546 [hep-th]].
%%CITATION = arXiv:1005.3546%%
}

%\LutyQC
\lref\LutyQC{
  M.~A.~Luty and R.~Rattazzi,
  ``Soft supersymmetry breaking in deformed moduli spaces, conformal theories, and N=2 Yang-Mills theory,''
JHEP {\bf 9911}, 001 (1999).
[hep-th/9908085].
%%CITATION = hep-th/9908085%%
}

%\StrasslerHT
\lref\StrasslerHT{
  M.~J.~Strassler,
  ``Nonsupersymmetric theories with light scalar fields and large hierarchies,''
[hep-th/0309122].
%%CITATION = hep-th/0309122%%
}

%\NelsonSN
\lref\NelsonSN{
  A.~E.~Nelson and M.~J.~Strassler,
  ``Suppressing flavor anarchy,''
JHEP {\bf 0009}, 030 (2000).
[hep-ph/0006251].
%%CITATION = hep-ph/0006251%%
}

%\NelsonMQ
\lref\NelsonMQ{
  A.~E.~Nelson and M.~J.~Strassler,
  ``Exact results for supersymmetric renormalization and the supersymmetric flavor problem,''
JHEP {\bf 0207}, 021 (2002).
[hep-ph/0104051].
%%CITATION = hep-ph/0104051%%
}

%\AzatovPS
\lref\AzatovPS{
  A.~Azatov, J.~Galloway and M.~A.~Luty,
  ``Superconformal Technicolor: Models and Phenomenology,''
Phys.\ Rev.\ D {\bf 85}, 015018 (2012).
[arXiv:1106.4815 [hep-ph]].
%%CITATION = arXiv:1106.4815%%
}

%\KutasovUX
\lref\KutasovUX{
  D.~Kutasov,
  ``New results on the \lq a theorem' in four-dimensional supersymmetric field theory,''
[hep-th/0312098].
%%CITATION = hep-th/0312098%%
}

%\PlehnPR
\lref\PlehnPR{
  T.~Plehn, M.~Spannowsky and M.~Takeuchi,
  ``Stop searches in 2012,''
[arXiv:1205.2696 [hep-ph]].
%%CITATION = arXiv:1205.2696%%
}

%\KutasovIY
\lref\KutasovIY{
  D.~Kutasov, A.~Parnachev and D.~A.~Sahakyan,
  ``Central charges and U(1)(R) symmetries in N=1 superYang-Mills,''
JHEP {\bf 0311}, 013 (2003).
[hep-th/0308071].
%%CITATION = hep-th/0308071%%
}

%\RattazziPE
\lref\RattazziPE{
  R.~Rattazzi, V.~S.~Rychkov, E.~Tonni and A.~Vichi,
  ``Bounding scalar operator dimensions in 4D CFT,''
JHEP {\bf 0812}, 031 (2008).
[arXiv:0807.0004 [hep-th]].
%%CITATION = arXiv:0807.0004%%
}

%\RattazziYC
\lref\RattazziYC{
  R.~Rattazzi, S.~Rychkov and A.~Vichi,
  ``Bounds in 4D Conformal Field Theories with Global Symmetry,''
J.\ Phys.\ A A {\bf 44}, 035402 (2011).
[arXiv:1009.5985 [hep-th]].
%%CITATION = arXiv:1009.5985%%
}

%\KarchQA
\lref\KarchQA{
  A.~Karch, T.~Kobayashi, J.~Kubo and G.~Zoupanos,
  ``Infrared behavior of softly broken SQCD and its dual,''
Phys.\ Lett.\ B {\bf 441}, 235 (1998).
[hep-th/9808178].
%%CITATION = hep-th/9808178%%
}

%\PolandWG
\lref\PolandWG{
  D.~Poland and D.~Simmons-Duffin,
  ``Bounds on 4D Conformal and Superconformal Field Theories,''
JHEP {\bf 1105}, 017 (2011).
[arXiv:1009.2087 [hep-th]].
%%CITATION = arXiv:1009.2087%%
}

%\BeniniMZ
\lref\BeniniMZ{
  F.~Benini, Y.~Tachikawa and B.~Wecht,
  ``Sicilian gauge theories and N=1 dualities,''
JHEP {\bf 1001}, 088 (2010).
[arXiv:0909.1327 [hep-th]].
%%CITATION = arXiv:0909.1327%%
}

%\KobayashiKZ
\lref\KobayashiKZ{
  T.~Kobayashi and H.~Terao,
  ``Sfermion masses in Nelson-Strassler type of models: SUSY standard models coupled with SCFTs,''
Phys.\ Rev.\ D {\bf 64}, 075003 (2001).
[hep-ph/0103028].
%%CITATION = hep-ph/0103028%%
}

%\VichiUX
\lref\VichiUX{
  A.~Vichi,
  ``Improved bounds for CFT's with global symmetries,''
JHEP {\bf 1201}, 162 (2012).
[arXiv:1106.4037 [hep-th]].
%%CITATION = arXiv:1106.4037%%
}

%\ChatrchyanTX
\lref\ChatrchyanTX{
  S.~Chatrchyan {\it et al.}  [CMS Collaboration],
  ``Combined results of searches for the standard model Higgs boson in $pp$ collisions at $\sqrt{s}=7$ TeV,''
Phys.\ Lett.\ B {\bf 710}, 26 (2012).
[arXiv:1202.1488 [hep-ex]].
%%CITATION = arXiv:1202.1488%%
}

%\PolandEY
\lref\PolandEY{
  D.~Poland, D.~Simmons-Duffin and A.~Vichi,
  ``Carving Out the Space of 4D CFTs,''
[arXiv:1109.5176 [hep-th]].
%%CITATION = arXiv:1109.5176%%
}

%\GreenNQ
\lref\GreenNQ{
  D.~Green and D.~Shih,
  ``Bounds on SCFTs from Conformal Perturbation Theory,''
[arXiv:1203.5129 [hep-th]].
%%CITATION = arXiv:1203.5129%%
}

%\KomargodskiPC
\lref\KomargodskiPC{
  Z.~Komargodski and N.~Seiberg,
  ``Comments on the Fayet-Iliopoulos Term in Field Theory and Supergravity,''
JHEP {\bf 0906}, 007 (2009).
[arXiv:0904.1159 [hep-th]].
%%CITATION = arXiv:0904.1159%%
}

%\
\lref\recent{
  C.~Brust, A.~Katz and R.~Sundrum,
  ``SUSY Stops at a Bump,''
[arXiv:1206.2353 [hep-ph]].
%%CITATION = arXiv:1206.2353%%
}

%\GohYR
\lref\GohYR{
  H.~-S.~Goh, M.~A.~Luty and S.~-P.~Ng,
  ``Supersymmetry without supersymmetry,''
JHEP {\bf 0501}, 040 (2005).
[hep-th/0309103].
%%CITATION = hep-th/0309103%%
}

%\AntoniadisEK
\lref\AntoniadisEK{
  I.~Antoniadis, J.~Iliopoulos and T.~Tomaras,
  ``On The Infrared Stability Of Gauge Theories,''
Nucl.\ Phys.\ B {\bf 227}, 447 (1983)..
%%CITATION = LPTENS 83/23%%
}

%\KaplanSK
\lref\KaplanSK{
  D.~B.~Kaplan,
  ``Dynamical Generation Of Supersymmetry,''
Phys.\ Lett.\ B {\bf 136}, 162 (1984)..
%%CITATION = HUTP-83/A078%%
}

\rightline{CERN-PH-TH/2012-156}
\Title{\vbox{\baselineskip12pt }} {\vbox{\centerline{Non-Perturbative Constraints on Light Sparticles}\centerline{from Properties of the RG Flow}}}
\smallskip
\centerline{Matthew Buican}
\smallskip
\bigskip
\centerline{{\it Department of Physics, CERN Theory Division, CH-1211 Geneva 23, Switzerland}} %
\vskip 1cm

\noindent
We study certain small supersymmetry-breaking perturbations of a large class of strongly coupled four-dimensional $R$-symmetric renormalization group (RG) flows between superconformal field theories in the ultraviolet (UV) and the infrared (IR). We analyze the conditions under which these perturbations scale to zero at leading order in the deep IR, resulting in accidental supersymmetry. Furthermore, we connect the emergence of IR supersymmetry in this context with a quantity that was recently conjectured to be larger at the UV starting points of the underlying supersymmetric flows than at the corresponding IR endpoints, and we propose a bound on emergent supersymmetry. Along the way, we prove a simple and useful non-perturbative theorem regarding the IR behavior of global flavor currents. Our results suggest general ways in which light stop particles can emerge and potentially influence physics at the Large Hadron Collider.

\bigskip
\Date{June 2012}

\newsec{Introduction}
From very simple principles, Quantum Field Theory (QFT) generates an astonishingly diverse set of phenomena. Gaining a quantitative handle on these phenomena is often very difficult, especially when non-perturbative effects become important. Given this state of affairs, a crucial challenge is to devise robust and calculable observables that obey simple and universal laws in order to constrain the possible behavior of large varieties of QFTs.

One particularly powerful set of constraints on QFT comes from studying observables in the deep UV and the deep IR of the renormalization group (RG) flow. In these limits, theories typically simplify and become conformal.\foot{Throughout this note we only discuss UV-complete QFTs. We should also note that while general principles dictate that the deep UV and the deep IR are scale invariant, it is an open question to find the precise conditions under which they are also conformal. In two dimensions scale invariance implies conformality under rather general assumptions \PolchinskiDY. In four dimensions, certain classes of scale-invariant theories are also known to be conformal \refs{\DorigoniRA, \AntoniadisGN, \LutyWW}, and it appears possible that this may be more generally true \LutyWW\ (note that there are certain known caveats for theories that have scalars with a shift symmetry or (dual) two-form fields; for a discussion of such cases in theories of the type we will discuss below, see \AntoniadisGN). For some results in $4-\epsilon$ dimensions, see \refs{\FortinKS, \NakayamaTK, \FortinIC}, and for a holographic perspective on this question, see \NakayamaWX.} As a simple example, one particularly natural quantity to compute in a conformal field theory (CFT) is the free energy of the theory compactified on a $D$-dimensional sphere, $F_{S^D}$. After appropriately regulating this quantity, one is left with a logarithmically divergent piece if $D$ is even and a finite piece if $D$ is odd. In four dimensions, the coefficient of the divergent piece is referred to as $a$, and is known to be smaller in the IR than the UV \refs{\KomargodskiVJ,\KomargodskiXV} (see also \refs{\ZamolodchikovGT\MyersTJ\CasiniKV\JafferisZI\MyersED\LiuEE\CasiniEI\AppelquistHR\ElvangST-\MaxfieldAW} for other interesting theorems governing the RG flow).

When the RG flow is supersymmetric one can often compute the above quantities much more easily and therefore enhance the power of the above theorems \refs{\AnselmiYS\IntriligatorJJ\KutasovUX\KutasovIY\JafferisUN-\ClossetVG}. Furthermore, when a theory is $R$ symmetric at all length scales (not just in the deep UV and the deep IR), it has more canonical conserved degrees of freedom, and so it is natural to imagine that there may be additional decreasing quantities along the RG flow.

Building on these ideas, we noted in \BuicanTY\ that since the RG-conserved $R$-current multiplet is often related to a local and gauge invariant long real multiplet, $U$,\foot{This relation occurs \KomargodskiRB\ whenever the theory also has a Ferrara-Zumino multiplet \FerraraPZ. Most theories of phenomenological interest have such multiplets with the exception of those that have field-independent Fayet-Iliopoulos terms \KomargodskiPC\ (sigma models with non-trivial target space topology also lack FZ multiplets \KomargodskiRB).} (containing the conformal anomaly) via \KomargodskiRB\
\eqn\rmult{
\bar D^{\dot\alpha}\CR_{\alpha\dot\alpha}=\bar D^2D_{\alpha}U~,
}
(where the $R$-current sits in the lowest component of $\CR_{\alpha\dot\alpha}$; the stress tensor and supercurrent sit in higher components of the multiplet) it is natural to consider the RG evolution of the two point function $\langle U_{\mu}(x)U_{\nu}(0)\rangle$, where $U_{\mu}\equiv-U|_{\theta\sigma^{\mu}\bar\theta}$. Note, however, that if the RG flow has conserved flavor (i.e., non-$R$) currents, $\hat J^a_{\mu}\equiv-\hat J^a|_{\theta\sigma^{\mu}\bar\theta}$ ($a=1,\cdot\cdot\cdot, N$),\foot{Conserved flavor currents sit in the $\theta\bar\theta$ components of real scalar superfields of dimension two, $\hat J_{a}$, that satisfy $\bar D^2\hat J_a=D^2\hat J_a=0$. We can pick out the conserved current component by acting on $\hat J_a$ with $\left[D_{\alpha},\bar D_{\dot\alpha}\right]$ and taking the bottom component. When a perturbative description in terms of chiral superfields exists, $\hat J_a$ is just a real bilinear in products of the chiral superfields with their conjugates weighted by the corresponding charges. For example, the baryon number current superfield in SQCD is of the form $J_B=Q\bar Q-\tilde Q\bar{\tilde Q}$.} then there is a family of $R$ symmetries (and corresponding $U$ operators) satisfying \rmult\ since $\CR^t_{\alpha\dot\alpha}=\CR_{\alpha\dot\alpha}+t_{a}{2\over3}\left[D_{\alpha},\bar D_{\dot\alpha}\right]\hat J^a$ is a conserved $R$ current for any real vector $t_a$ (with an associated operator $U^t=U+t_a\hat J^a$).\foot{The component stress tensor and supercurrent shift by improvement terms \KomargodskiRB.}

Therefore, in order to say something universal about the RG properties of $\langle U_{\mu}(x)U_{\nu}(0)\rangle$, we need a canonical way to define which $U$ (and therefore which $R$ symmetry) we would like to study. In \BuicanTY, we proposed studying the operator pair $(\CR_{\mu, {\rm vis}}, U_{\rm vis})$ defined by performing $a$-maximization \IntriligatorJJ\ in the deformed UV theory (the label \lq\lq vis" stands for \lq\lq visible" and is meant to remind us that the $R$ symmetry is a visible symmetry in the UV as opposed to an accidental one of the deep IR). More precisely, we imagine deforming the UV theory by adding a set of relevant operators to the Lagrangian (and / or turning on a set of vacuum expectation values (vevs)) that preserves an $R$ symmetry, $R_{\mu}^{(0), UV}$. If the deformed theory also preserves a set of flavor currents, $\hat J_{\mu, a}^{UV}$, then we fix their mixing with $U_{\mu}$ by considering the most general preserved $R$ current
\eqn\afunctional{
\CR^{t, UV}_{\mu}=\CR^{(0), UV}_{\mu}+\hat t^a\hat J_{\mu, a}^{UV}~,
}
and maximizing the anomaly functional
\eqn\amax{
\tilde a^{t, UV}=3\Tr\left(\CR^{t, UV}\right)^3-\Tr\CR^{t, UV}~,
}
i.e., we find the set of $\hat t^a=\hat t^{a}_*$ such that
\eqn\max{
\partial_{\hat t^a}\tilde a^{t, UV}|_{\hat t^a=\hat t^{a}_*}=0, \ \ \ \partial^2_{\hat t^a, \hat t^b}\tilde a^{t, UV}|_{\hat t^{a,b}=\hat t^{a,b}_*}<0~.
}
The resulting $R$ current and $U$ operator descend from conserved currents $(\CR_{\mu, {\rm vis}}, U_{\rm vis})$  of the UV SCFT.\foot{One caveat is that certain currents may have vanishing anomalies---the heuristic interpretation of such currents is that they act only on massive degrees of freedom. In order to fix the mixing of $U_{\mu, {\rm vis}}$ with such currents, $\tilde J^{UV}_{\nu, I}$, our prescription in \BuicanTY\ was to impose the orthogonality condition $\langle U^{UV}_{\mu, {\rm vis}}(x)\tilde J^{UV}_{\nu, I}(0)\rangle=0$. We will have more to say about such currents below. See \BuicanTY\ for discussions of additional subtleties.} These currents then flow to conserved currents $(\CR_{\mu, {\rm vis}}^{IR}, U_{\rm vis}^{IR})$ of the IR SCFT (although $U_{\rm vis}$ is {\it not} conserved in the bulk of the RG flow---its non-conservation measures the superconformal anomaly\foot{The crucial point is that even though $U_{\rm vis}$ is broken, it is related to the preserved $\CR_{\mu, {\rm vis}}$ current via an equation of the form \rmult, and so we can follow its RG evolution non-perturbatively. Typical long multiplets cannot be followed from the deep UV to the deep IR.}). In particular, the conserved vector, $U_{\mu, {\rm vis}}^{UV, IR}={3\over2}\left(\CR_{\mu, {\rm vis}}^{UV, IR}-\tilde R_{\mu}^{UV, IR}\right)$, measures the difference between the UV and IR limits of the RG-preserved $R$ symmetry current and the UV and IR superconformal $R$ currents, $\tilde\CR_{\mu}^{UV, IR}$, respectively.\foot{The superconformal $R$ current multiplets satisfy $\bar D^{\dot\alpha}\tilde \CR_{\alpha\dot\alpha}^{UV, IR}=0$, and they contain the traceless SCFT stress tensor and supercurrent in the higher components. The stress tensors and supercurrents in $\CR_{\mu, {\rm vis}}^{UV, IR}$ are generally related to the traceless ones by improvement transformations.} 

As a simple example of the above discussion, consider the case of SQCD (with $N_f<3N_c$); the above procedure yields charges
\eqn\UUVvisSQCD{
q_{\CR^{UV}_{\rm vis}}(Q)=q_{\CR^{UV}_{\rm vis}}(\tilde Q)=1-{N_c\over N_f}, \ \ \ q_{U^{UV}_{\rm vis}}(Q)=q_{U^{UV}_{\rm vis}}(\tilde Q)={1\over2}-{3N_c\over2N_f}~.
}
The IR limits of these expressions can also be easily computed \BuicanTY.

Given this discussion, we can then write the following two point functions in the UV and IR SCFTs
\eqn\consuUVIR{\eqalign{
\langle U^{UV}_{\mu, {\rm vis}}(x)U^{UV}_{\nu, {\rm vis}}(0)\rangle&={\tau_U^{UV}\over(2\pi)^4}\left(\partial^2\eta_{\mu\nu}-\partial_{\mu}\partial_{\nu}\right){1\over x^4}~, \cr\langle U^{IR}_{\mu, {\rm vis}}(x)U^{IR}_{\nu, {\rm vis}}(0)\rangle&={\tau_U^{IR}\over(2\pi)^4}\left(\partial^2\eta_{\mu\nu}-\partial_{\mu}\partial_{\nu}\right){1\over x^4}~,
}}
where the tensor structure of these two point functions is fixed by the conservation of $U_{\mu}$ up to two overall real coefficients, $\tau_U^{UV}\ne\tau_U^{IR}$.\foot{Note that since \rmult\ is invariant under holomorphic plus anti-holomorphic deformations in $U$, we should work modulo such deformations. One way to accomplish this is to simply study the piece in \consuUVIR\ that is proportional to $\eta_{\mu\nu}$. Furthermore, in the deep IR, $U$ often has a non-conserved holomorphic plus anti-holomorphic piece arising because of the emergence of Goldstone bosons \AbelWV. This non-conserved piece reflects the fact that the theory separates into an IR SCFT and a free Goldstone boson theory that is in a phase with a non-linearly realized shift symmetry.} From unitarity it immediately follows that $\tau_U^{UV}>0$. On the other hand, $\tau_U^{IR}\ge0$, with $\tau_U^{IR}>0$ if and only if $U_{\mu, {\rm vis}}^{IR}$ mixes with accidental symmetries of the IR fixed point. While it is then clear that $\tau_U^{UV}>\tau_U^{IR}$ if $U_{\rm vis}$ doesn't mix with accidental symmetries, in \BuicanTY\ we conjectured that, more generally, any mixing with accidental IR symmetries is bounded in the following way
\eqn\tauconj{
\tau_U^{UV}>\tau_U^{IR}~,
}
in all $R$-symmetric theories with an FZ multiplet. We checked this conjecture in a wide array of different theories with different dynamics and found no counterexamples \BuicanTY. 

While the above discussion may seem a bit abstract and removed from real-world particle physics, there are several reasons---some well-known and others not so well-known---to believe that this may actually not be the case:

\medskip\noindent
$\bullet$ First, the Large Hadron Collider (LHC) has observed a weakly-coupled Higgs at around 125 GeV. The most natural explanation for why such a state exists in the spectrum is that the Higgs is a particle in some as yet undiscovered SUSY Standard Model. Of course, if SUSY exists in nature, it is broken, and this breaking should occur in a separate \lq\lq hidden sector" module of particle physics \DimopoulosZB. On general grounds this hidden sector may well have an underlying $R$ symmetry (or at least an approximate one) \refs{\NelsonNF, \IntriligatorPY}. Furthermore, if SUSY is to be natural, its breaking should be dynamical (i.e., involving strong coupling), and so it is of great interest to derive laws governing the non-perturbative RG behavior of $R$-symmetric theories.

\medskip\noindent
$\bullet$ Second, it is often the case (and it is not unreasonable to conjecture that it is true more generally) that dynamical SUSY breaking (DSB) is accompanied by the emergence of accidental bosonic symmetries. One well-known example where this happens is the metastable SUSY-breaking model of Intriligator, Seiberg, and Shih \IntriligatorDD. There, the SUSY breaking occurs in the deep IR of mass-deformed SUSY QCD (SQCD) in the free magnetic range. In this regime, various accidental symmetries appear that rotate the IR magnetic degrees of freedom by phases. This observation leads one to believe that understanding bounds on emergent bosonic symmetries could help constrain theories of DSB. In fact, in \BuicanTY\ we showed how this idea could work in the case of the $SU(2)$ gauge theory of Intriligator, Seiberg, and Shenker \IntriligatorRX.

\medskip\noindent
$\bullet$ Third, the fact that SUSY hasn't yet been observed has placed strong lower bounds on first generation squark masses at around 1 TeV \refs{\ATLASAE, \ChatrchyanTX}. If SUSY is to remain natural, we are forced to consider third generation squarks with lower relative masses, and we are pushed in the direction of the types of models considered recently in \refs{\SundrumGV\BarbieriPD\BarbieriAR\CraigYK\GherghettaWC\JeongEN\EssigQG\KatsQH\PapucciWY\BrustTB\DelgadoKR\DesaiTH\AkulaJX\AjaibHS\IshiwataAB\LodoneAA\HeTP\ArvanitakiCK\AuzziEU\CsakiFH\CraigYD\LarsenRQ\CraigHC\CraigDI\PlehnPR\AlvesFT\HanFW\KaplanGD-\recent} (and early on in \refs{\DineNP, \DimopoulosMI}; see \CraigYD\ for a list of references from this era). This hierarchical squark flavor structure implies that the stop / top (broken) multiplet is a sector of quasi-emergent SUSY.
\medskip

The final comment above will be particularly interesting for us in what follows. Indeed, as we will explain in the next section, we can translate---in a limited but precise sense---\tauconj\ into a lower bound on the amount of accidental SUSY (and therefore, if a perturbative description exists in the IR, an upper bound on certain leading order scalar masses) in a large class of non-perturbative RG flows.\foot{For previous work on formal aspects of accidental SUSY in various different contexts, see for example \refs{\AntoniadisEK\KaplanSK\GohYR-\StrasslerHT, \SundrumGV}.}

Furthermore, by studying theories that satisfy the stronger inequality $\tau_U^{UV}>0=\tau_U^{IR}$ (i.e., those theories for which accidental symmetries do not mix with the IR superconformal $R$ current), we can make contact with the recently advocated approach in  \CsakiFH\ for producing light stop particles using a small SUSY-breaking deformation of SQCD at the boundary of the conformal window (i.e., $N_f=3N_c/2$). We will apply our results to study various model-independent aspects and extensions of this idea. We should also note, however, that there are at least two constraints on generating light scalar masses directly in SQCD with $N_f=3N_c/2$: the IR suppression of scalar masses is at most logarithmic (although it is possible that one-loop factors may make this suppression sufficient when embedding this construction in a phenomenological setting) and generic soft masses in the UV give rise to unsuppressed (and tachyonic) soft masses in the IR (we discuss how to forbid such soft terms at length). 

The plan of this paper is as follows: after illustrating the connection between \tauconj\ and non-perturbative lower bounds on emergent SUSY as well as making contact with \CsakiFH\ in the next section, we step back and consider how universal emergent SUSY is within the class of RG flows we study. To answer this question it turns out to be critical to understand how RG-preserved flavor currents behave non-perturbatively. With this goal in mind, we prove a simple, quantum mechanical theorem on the IR behavior of RG-preserved flavor currents and comment on its implications. We then discuss a corollary for asymptotically free theories (relegating the proof to the Appendix). Finally, we conclude with a discussion of constraints on emergent SUSY and light sparticles in QFT with an eye towards future phenomenological applications.

\newsec{Implications of $\delta\tau_U>0$ on accidental SUSY}
As we discussed in the introduction, $U_{\rm vis}^{UV}$ is a unique and well-defined operator multiplet that exists in every $R$-symmetric theory with an FZ multiplet (see \BuicanTY\ for detailed computations involving this object in many examples). Its bottom component also provides a canonical way to break SUSY in such RG flows. Indeed, consider deforming the UV SCFT by turning on the following deformation
\eqn\defUV{
\delta S_{UV}|_{SSB}=-\int d^4x\ \lambda\cdot U^{UV}_{\rm vis}|~.
}
Here $\lambda$ is a coupling of dimension two, and $U^{UV}_{\rm vis}|$ is the bottom component of the $U^{UV}_{\rm vis}$ multiplet (the label \lq\lq SSB" stands for \lq\lq SUSY breaking"). In general, the UV theory may be a strongly interacting SCFT with no Lagrangian description (see \BeniniMZ\ for some tools to analyze such theories). In this case, $U^{UV}_{\rm vis}$ does not admit an interpretation in terms of particles. Instead, it is simply a multiplet of dimension two, and adding the bottom component to the action, as in \defUV, corresponds to (softly) breaking SUSY.

If the theory in question is asymptotically free and has matter superfields, $\Phi_i$, we find that\foot{We suppress factors of gauge superfields that are required for gauge invariance.}
\eqn\UUV{
U^{UV}_{\rm vis}=-\sum_iq_{U^{UV}_{\rm vis}}(\Phi_i)\cdot\Phi_i\bar\Phi^i, \ \ \ q_{U^{UV}_{\rm vis}}(\Phi_i)={3\over2}\left(q_{\CR^{UV}_{\rm vis}}(\Phi_i)-{2\over3}\right)~,
}
where $q_{U^{UV}_{\rm vis}}(\Phi_i)$ and $q_{\CR^{UV}_{\rm vis}}(\Phi_i)$ are the charges assigned to $\Phi_i$ under the corresponding symmetries. Therefore, in this case, \defUV\ can be interpreted as a soft mass for the scalars
\eqn\pertdef{
\delta S_{UV}|_{SSB}=-\int d^4x \ (m^{UV}_i)^2\cdot  \bar\phi^i\phi_i, \ \ \ (m^{UV}_i)^2\equiv m^2\cdot q_{U^{UV}_{\rm vis}}(\phi_i)~,
}
where $\phi_i\equiv\Phi_i|$.

In order to maintain control over our theory, we work in the probe SUSY breaking approximation:\foot{Various authors have studied fundamental aspects of such softly-broken RG flows (early and closely related works include \refs{\ArkaniHamedWC, \LutyQC}---see also \KarchQA\ for the first results on soft terms in the conformal window of SQCD; our approach follows the formalism in \AbelWV; other works exploring related topics include \refs{\SundrumGV, \StrasslerHT, \NelsonSN, \KobayashiKZ, \NelsonMQ}).} the limit in which the SUSY-breaking term doesn't back-react on the strong dynamics of the RG flow. In other words, we deform the UV theory by
\eqn\deltas{
\delta S_{UV}=-\int d^4x\left(\int d^2\theta\ \lambda_{\CO}\cdot\CO+{\rm h.c.}+\lambda\cdot U_{\rm vis}^{UV}|\right)~,
}
where $\CO$ is an $R$ symmetry-preserving superconformal primary of dimension $d_{\CO}\le3$,\foot{If $d_{\CO}=3$, we assume that $\lambda_{\CO}$ is marginally relevant and therefore corresponds to turning on an asymptotically free gauge coupling. Furthermore, in writing \deltas, we have used the fact that unitarity bounds rule out relevant K\"ahler deformations.} \foot{We could also imagine turning on a SUSY and $R$-symmetric vev, $\langle\hat\CO\rangle$, for some lorentz scalar of dimension $d_{\hat\CO}>1$ (if $d_{\hat\CO}=1$, $\hat\CO$ is a free field and turning on the vev does not initiate an RG flow). If this vev is the dominant supersymmetric breaking of conformality, we should assume that $|\langle\hat\CO\rangle|^{2/d_{\hat\CO}}\gg|\lambda|$ parametrically in order to work in the probe approximation.} and we assume the following parametric inequalities
\eqn\probedef{\eqalign{
\left|\lambda_{\CO}\right|^{2\over 3-d_{\CO}}&\gg|\lambda|, \ \ \ d_{\CO}<3~,\cr \Lambda^2&\gg|\lambda|, \ \ \ d_{\CO}=3~,
}}
where $\Lambda$ is the dynamical scale induced by turning on the marginally relevant $\CO$ (in the case of asymptotically free theories---as opposed to a generic interacting UV theory for which we turn on an asymptotically free coupling---the second line in \probedef\ corresponds to turning on parametrically small soft masses, $|m|^2\ll\Lambda^2$ in \pertdef).\foot{In the case of multiple relevant SUSY deformations in the UV, $\delta W_{UV}=-\lambda_{\CO_i}\cdot\CO_i$, we assume that \probedef\ applies for each $\lambda_{\CO_i}$ and $\Lambda_i$.}

Following the RG flow into the deep IR, we find that in our approximation the SUSY breaking deformation flows to \refs{\AbelWV, \ArkaniHamedWC, \LutyQC}
\eqn\defIR{
\delta S_{IR}|_{SSB}=-\int d^4x\ \lambda\cdot U^{IR}_{\rm vis}|~.
}
Let us now define $\delta \hat S_{IR}|_{SSB}$ to be equal to \defIR\ modulo chiral plus anti chiral terms (recall from the discussion above that we define $\tau_U$ by following $U_{\rm vis}$ modulo holomorphic plus anti-holomorphic terms; in addition, such terms often simply vanish on symmetry grounds \AbelWV). From the conjecture $\tau_U^{UV}>\tau_U^{IR}$, we see that the SUSY breaking operator appearing in the IR has a smaller norm (i.e., two-point function) than the SUSY breaking operator appearing in the UV. This inequality constitutes an upper bound on the amount of IR versus UV SUSY breaking (we will from now on assume that any holomorphic plus anti-holomorphic terms are absent; they will not affect our conclusions and can be appropriately reincorporated if necessary).\foot{Note that for generic UV soft terms we {\it do not} expect such behavior. Indeed, for generic $U^{UV}\ne U_{\rm vis}^{UV}$ operators there is no decreasing behavior of the corresponding two-point function in the deep IR \BuicanTY.}

When the theory is free in the IR, we find that 
\eqn\UIR{
U^{IR}_{\rm vis}=-\sum_aq_{U^{IR}_{\rm vis}}(\hat\Phi_a)\cdot\hat\Phi_a\bar{\hat\Phi}^a, \ \ \ q_{U^{IR}_{\rm vis}}(\hat\Phi_a)={3\over2}\left(q_{\CR^{IR}_{\rm vis}}(\hat\Phi_a)-{2\over3}\right)~,
}
where the $\hat\Phi_a$ are the emergent degrees of freedom. Therefore, we see in this case that
\eqn\deltaSIRpert{
\delta S_{IR}|_{SSB}=-\int d^4x \ (m^{IR}_a)^2\cdot  \bar{\hat \phi}^a\hat \phi_a, \ \ \ (m^{IR}_a)^2\equiv m^2\cdot U^{IR}_{\rm vis}(\hat\phi_a)~,
}
where $\hat\phi_a=\hat\Phi_a|$. If the UV theory is asymptotically free (as in \pertdef), the bound on IR SUSY breaking discussed above then amounts to an inequality on the sum of the fourth power of the soft masses
\eqn\susybrpert{
\sum_a (m_a^{IR})^4<\sum_i(m_i^{UV})^4~,
}
and so this clearly constitutes an upper bound on the amount of SUSY breaking in the IR (and hence a lower bound on the degree to which SUSY emerges in the IR). Let us stress again that this inequality {\it does not} apply for general UV and IR soft terms different from $U_{\rm vis}|$.\foot{We would also like to emphasize that we have derived the inequality in \susybrpert\ from $\delta\tau_U>0$ based on the assumption that the SUSY breaking deformation is parametrically smaller than the (marginally) relevant SUSY deformations. Take $SU(N_c)$ SQCD with $N_c+2<N_f\le 3N_c/2$ flavors as an example. Let us consider starting in the UV, turning on the $SU(N_c)$ gauge coupling, and adding masses $\delta W_{UV}=m_iQ_i\tilde Q_i$ for some subset of the flavors ($i=n, ..., N_f$, with $n>N_c+2$). Whether the masses are small (compared to the dynamical scale, $\Lambda$) or not, the deep IR theory is the free $SU(n-1-N_c)$ magnetic SQCD with $n-1$ flavors, and the flow satisfies $\delta\tau_U>0$, where, by definition, we include all the relevant deformations in the UV in determining $U_{\rm vis}$ and hence $\tau_U$. In this case, provided the SUSY breaking is parametrically small (i.e., $m^{UV}\ll m_i, \Lambda$, where $m^{UV}$ is the characteristic UV soft SUSY-breaking scale), one can check that \susybrpert\ is indeed satisfied in the deep IR. Note, however, that if $m_i\ll\Lambda$, but $m^{UV}$ is not parametrically smaller than the $m_i$, it may happen that $\hat m_i\ll m^{IR}\ll\Lambda$ (here $\hat m_i$ are the magnetic masses that are induced via Higgsing by the dual of the $m_i$ deformations in the IR theory). In this case, at the scale $m^{IR}$, $U_{\rm vis}$ is an operator in $SU(N_f-N_c)$ magnetic SQCD with $N_f$ flavors, and it can happen that the inequality in \susybrpert\ does not hold. This discussion does not contradict the statement that $\delta\tau_U>0$ since this inequality is defined in terms of quantities at the UV and IR endpoints of the flow. Indeed, in the SUSY theory, the $\hat m_i$ masses eventually drive the theory to the $SU(n-1-N_c)$ magnetic SQCD endpoint, and $\delta\tau_U>0$. However, if we are interested in the limit where $\hat m_i\ll m^{IR}\ll\Lambda$, we can find a $U_{\rm vis}$ SUSY-breaking deformation satisfying \susybrpert\ at the scale $m^{IR}$ simply by treating $\delta W_{UV}=m_iQ_i\tilde Q_i$ as a small perturbation and adding the soft term proportional to the $U_{\rm vis}^{UV}|$ operator of the massless theory.}

At this point, several comments are in order:

\medskip\noindent
$\bullet$ If $U^{IR}_{\rm vis}=0$, this is in some sense the most extreme manifestation of $\tau_U^{UV}>\tau_U^{IR}$, and we say that SUSY is emergent since the IR SUSY breaking is small. Note that $U_{\rm vis}^{IR}=0$ if and only if accidental symmetries do not mix with the IR superconformal $R$ current. However, while $U^{IR}_{\rm vis}=0$ has a precise meaning in the underlying theory, we should not take this equation to mean that the SUSY breaking becomes unimportant in the IR. Indeed, the degree to which SUSY is emergent depends on whether $U_{\rm vis}$ flows to zero logarithmically or as a power law (our discussion here closely follows the discussion in \AbelWV). Note that $U_{\rm vis}$ flows to zero logarithmically only if the approach to the IR fixed point is via a marginally irrelevant operator. From the general results of \GreenDA, an operator is marginally irrelevant only if it breaks an (accidental) symmetry, $J$, of the IR SCFT. Therefore, $U_{\rm vis}$ flows to zero logarithmically only if it mixes with an accidental symmetry of the IR SCFT away from the fixed point.\foot{In other words, we have $U_{\rm vis}=\gamma\cdot J$, where $J$ is an accidental symmetry of the fixed point and $\gamma\to0$ logarithmically in the deep IR. This mixing can be computed in conformal perturbation theory using the general techniques in \GreenDA.} This is precisely what happens in SQCD for $N_f=3N_c/2$ with its plethora of accidental symmetries. In this case, $U_{\rm vis}^{UV}|={1\over2}\left(Q\bar Q+\tilde Q\bar{\tilde Q}\right)$ and so \pertdef\ becomes
\eqn\SUVSQCDbd{
\delta S_{UV}|_{SSB}=\int d^4x\ {m^2\over2}\left(Q\bar Q+\tilde Q\bar{\tilde Q}\right)~,
}
This deformation results in emergent SUSY in the IR in the sense that $Q\bar Q+\tilde Q\bar{\tilde Q}\to0$ \refs{\ArkaniHamedWC,\LutyQC,\AbelWV}. However, since this flow is logarithmic, the SUSY breaking deformation enters the dynamics suppressed only logarithmically.\foot{In this case, we don't need the general results of \GreenDA. Let us consider a slightly simpler example that illustrates the same point---generalizing to SQCD with $N_f=3N_c/2$ is straightforward. Suppose our underlying SUSY theory of the deep IR consists of a single chiral superfield, $\Phi$, with $\CL=\int d^4\theta Z(\mu)\cdot\bar\Phi\Phi+\left(\int d^2\theta\lambda\Phi^3+{\rm h.c.}\right)$, where $\mu$ is the RG scale and $Z$ is the scale-dependent wave-function factor. Working in a holomorphic renormalization scheme, we can compute $U_{\rm vis}\sim{d \over d\log\mu}K$, where $K$ is the K\"ahler potential. Clearly $U_{\rm vis}\sim|\lambda|^2\bar\Phi\Phi$ (up to one-loop factors), where the (physical) $\lambda$ (and hence $U_{\rm vis}$) flows to zero in the deep IR logarithmically---note that at the IR fixed point $\bar\Phi\Phi$ is just the current superfield corresponding to the accidental symmetry that rotates $\Phi$ by a phase. Now, if we add the SUSY breaking deformation proportional to $U_{\rm vis}$ into the theory, we see that the resulting masses are only suppressed by the logarithmic running of the coupling.}
On the other hand, if the IR has no accidental symmetries, then we see that $U_{\rm vis}$ flows to zero as a power law since it must mix with some leading IR SCFT operator of dimension $d>2$ and so the UV deformation enters the IR dynamics suppressed by $\sim{m^2\over\Lambda^{d-2}}$ (if $d>4$, then the SUSY breaking term is irrelevant and flows exactly to zero in the deep IR).
This is precisely the situation in the conformal window of SQCD (i.e., for $3N_c/2<N_f<3N_c$). Let us note that logarithmic behavior can also occur in the IR for an interacting SCFT with accidental symmetries, and the power law behavior can occur in a theory that flows to a free fixed point.\foot{For example, it may happen that the deep IR of the underlying SUSY theory looks like $\CL_{IR}=\int d^4\theta\left(\bar{\hat\Phi_a}\hat\Phi^a+\kappa\left(\bar{\hat{\Phi}_a}\hat\Phi^a\right)^2\right)$,  where appropriate symmetries keep there from being marginally irrelevant operators constructed out of the weakly coupled emergent $\hat\Phi_a$. If these symmetries are symmetries of the full theory, then $U_{\rm vis}^{IR}\to0$ as a power law.}

\medskip\noindent
$\bullet$ The authors in \CsakiFH\ considered embedding stops and Higgses as emergent degrees of freedom in SQCD with $N_f=3N_c/2$. They added precisely the deformation \SUVSQCDbd\ to the UV Lagrangian.\foot{In their phenomenological construction, the authors of \CsakiFH\ also considered adding various relevant deformations (and Yukawa terms which become relevant at finite gauge coupling) to the UV theory in order to eliminate exotics and generate the symmetry-breaking required for phenomenology. By applying the results of \refs{\ArkaniHamedWC,\LutyQC,\AbelWV}, these deformations are implicitly considered as perturbations of the underlying strong dynamics.}
Treating the first two generations of matter as SQCD singlet spectators that receive UV soft masses comparable to those of the UV \lq\lq ancestors" of the third generation (and Higgs sector), the IR soft masses of the third generation (and Higgs sector) are then suppressed (logarithmically) with respect to the masses of the first two by the SQCD dynamics (one must also ensure that the transition scale between the IR and UV dynamics is sufficiently low so that perturbative SSM corrections to the Higgs and third generation scalar masses don't destabilize the pattern of soft masses arising from the strong dynamics). We can now describe the basic idea the authors advocated in our language: assuming the existence of a conformal window with no accidental symmetries (as in SQCD), a soft mass related to $U_{\rm vis}^{UV}$ flows to zero in the deep IR as a power law. Assuming there is a non-trivial boundary to the conformal window where a calculable free theory emerges (as in SQCD), it is natural to assume by continuity that $U_{\rm vis}^{UV}$ flows to zero there. Thinking along these lines, it might also be of interest to try to find such a boundary where the power law suppression remains even in the free theory (although this is somewhat in tension with the necessity of having non-trivial Yukawa couplings).

\medskip\noindent
$\bullet$ The term in \defUV\ is not the most general explicit SUSY-breaking deformation we can add. For example, the UV SCFT may contain a whole set of global (non-$R$) currents, and we can then consider adding their bottom components to the UV theory as well. Let us label these currents as follows: $\hat J_a^{UV}$, $J_A^{UV}$, and $U^{UV}_{\rm vis}$. The index $a$ runs over the currents that are preserved along the full RG flow, while $U^{UV}_{\rm vis}$ is the operator we've introduced above, and the index $A$ runs over the remaining flavor currents that are broken away from the UV SCFT. Given this set of operators, we can consider deforming the UV theory as follows
\eqn\gencurrUV{
\delta S_{UV}|_{SSB}=-\int d^4x\ \left(\lambda_U\cdot U^{UV}_{\rm vis}|+\lambda_a\cdot \hat J^{UV}_a|+\lambda_A\cdot J_A^{UV}|\right)~.
}
In general we can turn on many additional types of relevant SUSY breaking deformations, but we will limit ourselves in what follows to studying general $R$-symmetric RG flows deformed as in \gencurrUV. The heuristic reason for this is just that such operators include scalar soft terms (and their non-perturbative generalizations), and the more limited scope of such theories will allow us to make more general statements.\foot{Another important SUSY-breaking deformation of phenomenological interest is the gaugino mass. Provided this deformation is parametrically small, we can follow it at linear order \refs{\ArkaniHamedWC,\LutyQC,\AbelWV} even for strongly coupled gauge groups. We should note that masses for SSM gauginos (these fields often arise from weakly gauging a global symmetry of the underlying strongly-coupled theory) may furnish important perturbative corrections to the emergent third generation scalar and Higgs sector masses. We will neglect all such contributions below, because we will assume a low crossover scale from the IR to UV dynamics. However, it is straightforward to add gaugino masses into our theories and analyze their effects.}

\medskip\noindent
$\bullet$ Given the set of deformations in \gencurrUV, we would like to understand how universal emergent SUSY is in the IR---that is, what conditions we should impose upon $\delta S_{UV}|_{SSB}$ so that $\delta S_{IR}|_{SSB}\to0$ at leading order. As we will see in the next section, we can place particularly strong constraints on the behavior of RG-preserved global symmetry currents, $\hat J_a$, and hence soft terms proportional to their lowest components. To understand the importance of these terms, we can again consider SQCD with $N_f\le 3N_c/2$. Adding a more general soft term in the UV of the form
\eqn\gencurrUVx{
\delta S_{UV}|_{SSB}=\int d^4x\ \left({m^2\over2}\left(Q\bar Q+\tilde Q\bar{\tilde Q}\right)+m'^2\left(Q\bar Q-\tilde Q\bar{\tilde Q}\right)\right)~,
}
results in unsuppressed (and tachyonic) masses in the IR at leading order in $m'$ since \AbelWV\ (see also the discussion in \refs{\LutyQC, \AzatovPS}) 
\eqn\gencurrUVxx{
\delta S_{IR}|_{SSB}=\int d^4x\ m'^2\left({N_c\over N_f-N_c}\right)\left(q\bar q-\tilde q\bar{\tilde q}\right)+\cdot\cdot\cdot~.
}
Here the $q$ and $\tilde q$ are the magnetic squarks. This result follows from the SUSY mapping \SeibergPQ\ of the bottom component of the baryon current superfield, $J_B^{UV}|=Q\bar Q-\tilde Q\bar{\tilde Q}\to {N_c\over N_f-N_c}\left(q\bar q-\tilde q\bar{\tilde q}\right)=J^{IR}_B|$ (similar results apply for more general linear combinations of UV soft terms that include terms related to the bottom components of the global $SU(N_f)_L\times SU(N_f)_R$ current superfields). The fact that the baryon current does not decouple and that therefore the soft term propotional to its bottom component remains at leading order in the IR is a simple example of a more general behavior we will find in the next section (similar points have been raised in \refs{\NelsonMQ,\SundrumGV}; however, our approach will apply to all the theories we will study, and it will follow simply from unitarity and 't Hooft anomaly matching).

\newsec{A theorem on the IR behavior of RG-preserved currents}
In this section we will prove a theorem concerning the IR behavior of RG-preserved flavor current superfielields, $\hat J_a$, with a view toward understanding the RG evolution of soft term deformations proportional to the bottom components of these operators.

Since we would like to get a handle on the broader question of when the soft terms in \gencurrUV\ flow to zero at leading order in the IR, we should study theories in which $U^{IR}_{\rm vis}=0$. As discussed in the previous section, this assumption amounts to studying the theories for which the infrared superconformal $R$ current, $\tilde R_{\mu}^{IR}$, descends from a preserved $R$-current, $\CR_{\mu, {\rm vis}}$, of the full RG flow, i.e.
\eqn\rvisIR{
\tilde R_{\mu}^{IR}=\lim_{E\to0}\CR_{\mu, {\rm vis}}~,
}
where $E$ is the RG scale.

Before proceeding to the theorem, let us note that the proof does not depend on the existence of an FZ multiplet or therefore a well-defined $\left(\CR_{\mu, {\rm vis}}, U_{\rm vis}\right)$ operator pair. As a result, we will prove the theorem under the weaker assumption that
\eqn\rvisIRi{
\tilde R_{\mu}^{IR}=\lim_{E\to0}\CR_{\mu}~,
}
for some general RG-preserved $R$ current satisfying $\bar D^{\dot\alpha}\CR_{\alpha\dot\alpha}=\chi_{\alpha}$ with $\bar D_{\dot\alpha}\chi_{\alpha}=D^{\alpha}\chi_{\alpha}-\bar D_{\dot\alpha}\bar\chi^{\dot\alpha}=0$. However, when we apply this theorem to theories with SUSY-breaking, we will continue to assume the existence of an FZ multiplet and an $\left(\CR_{\mu, {\rm vis}}, U_{\rm vis}\right)$ operator pair so that we can make contact with soft masses.

Under such conditions we can state a simple theorem:
\medskip
\noindent
{\bf Theorem:} A necessary and sufficient condition for the RG-preserved non-$R$ currents, $\hat J_a$, to flow to zero in the deep IR is that all the 't Hooft anomalies involving these currents vanish, i.e.
\eqn\tHooft{
\Tr \hat J_a\hat J_b\hat J_c=0, \ \ \ \Tr R\hat J_a\hat J_b=0, \ \ \ \Tr R^2\hat J_a=0, \ \ \ \Tr \hat J_a=0~.
}

\medskip
\noindent
{\bf Proof:} Necessity is trivial and follows from the following observation: if one of these anomalies is non-zero, 't Hooft anomaly matching forces there to be light fields charged under the corresponding symmetries. Sufficiency follows from the following reasoning. First, recall from \rvisIRi\ that $R_{\mu}$ flows to the IR superconformal $R$ current, $\tilde R_{\mu}^{IR}$. Let us then suppose that $\hat J_a\to \hat J_a^{IR}\ne0$. In this case, $\Tr \tilde R_{IR}\hat J_a\hat J_a<0$ (by unitarity) and so we must have $\Tr R\hat J_a\hat J_a<0$. This inequality conflicts with the second equation in \tHooft, and so it must be the case that $\hat J_a\to0$ in the IR. {\bf q.e.d.}
\medskip

This theorem holds for RG flows with and without Lagrangian descriptions in the UV and the IR. Note that the necessary part of the above theorem is always true (and somewhat trivial), while the sufficient part is special to the above class of theories. Also note that the flavor currents and the $R$ current are on a very different footing as far as their IR behavior is concerned. Indeed, if it were the case that $R\to0$, then the IR theory would be trivial (it would not have a stress tensor; heuristically this corresponds to giving mass to all the degrees of freedom in the QFT; in our class of theories this happens if and only if $\Tr R=\Tr R^3=0$). On the other hand, $\hat J_a\to0$ does not imply that the IR theory is trivial.\foot{Indeed, we can often make the sub-sector on which a given current, $\hat J_a$, acts massive without rendering the full theory trivial. Consider for example SQCD with $N_f=3N_c-1$. Adding a large mass, $m\gg\Lambda$, for one flavor, $\delta W_{\rm UV}=mQ_1\tilde Q_1$, does not render the IR theory trivial, but it does render the current $\hat J_{11}$ for the symmetry that acts on $Q_1$ and $\tilde Q_1$ with opposite phases (leaving the other quarks invariant) massive; in particular, $\hat J_{11}\to0$.} However, we will now see that the above theorem implies that the $\hat J_a$ currents cannot flow to zero in a broad class of theories. 

\subsec{Asymptotically free theories}
In this section we specialize to asymptotically free theories and study some of the consequences of the theorem  (we emphasize once more, however, that the theorem applies to interacting UV theories as well). Let us first state a corollary: 

\medskip\noindent
{\bf Corollary}: Consider the set of asymptotically free theories with simple gauge group and vanishing superpotential such that \rvisIRi\ holds. If such a theory has a set of non-anomalous global symmetry currents, $\hat J_a$, then it follows that some of these currents do not flow to zero in the IR.

\medskip\noindent
{\bf Sketch of proof:} We prove by contradiction. Using the theorem we should try to impose that the 't Hooft anomalies in \tHooft\ all vanish. A simple exercise in linear algebra reveals that this is inconsistent with the fact that our theory has a non-anomalous $R$ symmetry. For the interested reader, we relegate the full proof to the Appendix.

\medskip
Note that this proof applies whether the theory is IR free or not. For example, SQCD with $3N_c/2\le N_f<3N_c$ is covered by the above corollary. Indeed, as noted in the previous section, the baryon current does not decouple in the IR (in fact neither do the $SU(N_f)_L\times SU(N_f)_R$ currents).

It is also clear that if we turn on a superpotential, $\delta W$, that renders some of the currents massive (or breaks them explicitly) or if we consider gauging some of the flavor symmetries, the above corollary guarantees that, as long as the deformations are small enough, the corresponding currents remain important in the IR. More generally, however, even when we turn on large deformations, the theorem provides powerful constraints on the decoupling of any global symmetries that remain (we will return to this point in the next section when we discuss the RG flow of the corresponding soft terms).

\newsec{Comments on the universality of accidental SUSY}
In this section we would like to use the results from the previous sections to analyze the universality of emergent leading-order SUSY in the class of RG flows we have defined above. We conclude by describing some criteria that one might use to try to engineer natural theories of light sparticles.

To that end, consider again the UV SUSY breaking deformation in \gencurrUV, which we reproduce below for ease of reference
\eqn\gencurrUVii{
\delta S_{UV}|_{SSB}=-\int d^4x\ \left(\lambda_U\cdot U^{UV}_{\rm vis}|+\lambda_a\cdot \hat J^{UV}_a|+\lambda_A\cdot J_A^{UV}|\right)~.
}
In the deep IR, at leading order in SUSY breaking, this deformation flows according to the operator mappings in the underlying SUSY theory \refs{\AbelWV, \ArkaniHamedWC, \LutyQC}
\eqn\gencurrUViii{
\delta S_{IR}|_{SSB}=-\int d^4x\ \left(\lambda_U\cdot U^{IR}_{\rm vis}|+\lambda_a\cdot \hat J^{IR}_a|+\lambda_A\cdot J_A^{IR}|\right)~.
}
Note, however, that the mappings for the $J_A$ operators (recall these currents are broken away from the UV SCFT and are not directly related to the unbroken $R$ current) are generally unknown even at leading order \AbelWV. The basic reason for this incalculability is that unlike $U$, such broken currents are not related to a manifestly preserved RG quantity. This incalculability will play a role in our engineering strategy.

Combining this discussion with the theorem and the corollary we conclude:

\medskip\noindent
$\bullet$ In general theories of the type we are interested in (i.e., with $U^{IR}_{\rm vis}=0$), soft masses proportional to the bottom components of RG-preserved currents, $\hat J_a|$, flow to zero at leading order if and only if the corresponding current has vanishing 't Hooft anomalies. This requirement is highly restrictive. Furthermore, such soft terms can grow in the IR since the central functions of unbroken currents often get larger at long distances (consider, for example, the central values of the baryon current in SQCD in the free magnetic phase).

\medskip\noindent
$\bullet$ Asymptotically free theories with vanishing superpotentials and simple gauge groups will have leading-order soft terms in the IR unless we restrict the set of UV soft terms to a subset of measure zero. The potential loophole is theories with a single representation of multiplicity one. To understand this statement, suppose $U^{IR}_{\rm vis}=0$, and recall that asymptotically free theories of the type we have described are specified by the following data: a (simple) gauge group, $G$, and representations, $r_i$, of multiplicities, $n_i$ ($i=1,..., s$). The resulting anomaly-free global symmetry group is then
\eqn\globalsymmpre{
S=U(1)_R\times U(1)^{s-1}\times\prod_{i=1}^sSU(n_i)~.
}
Clearly there will be flavor symmetry unless $s=n_1=1$. If this is not the case, the corresponding coefficients for the preserved currents in \gencurrUVii\ should be set to zero in order to ensure suppressed IR soft terms.

\medskip\noindent
$\bullet$ Small deformations of such theories do not change these conclusions since we can expand in the deformation parameters.

\subsec{Consequences for models of light sparticles: UV aspects}
The basic idea for making contact with phenomenology proceeds along the lines of \CsakiFH. We imagine that we want to produce light emergent states of phenomenological interest. The \lq\lq ancestors" of these states receive UV soft masses that should be written in terms of the $U_{\rm vis}^{UV}$ multiplet (in order for the emergent masses to be suppressed in the IR) of some strongly coupled RG flow (let us emphasize again that we will assume that the transition scale from the UV to the IR dynamics is sufficiently low so that perturbative corrections from SSM fields will not destabilize any hierarchy we obtain from the strongly coupled flow). The remaining matter fields receive comparable UV soft masses but are singlets under the strong dynamics.

The above discussion suggests that in order to engineer a natural, UV-insensitive, model of emergent SUSY, we would like to consider theories in which we are forbidden from adding UV soft terms proportional to unbroken currents (related comments have been made by many other authors including, but not limited to, those in \NelsonMQ, \SundrumGV, and \AzatovPS) and in which, for calculability, we limit the number of UV SCFT symmetries that we break along the RG flow. Clearly, we would also like to consider theories in which $U_{\rm vis}^{IR}=0$.

We can address this problem in several ways. For example, we can consider imposing discrete symmetries on the UV soft terms, turning on gauge couplings, deforming the theory by additional relevant superpotential terms in order to break some of the symmetries (note that this helps with limiting the number of RG-preserved currents, but, as discussed above, it can make our model less calculable), or simply trying to specify an SCFT with the minimal symmetry content required for phenomenology. Let us consider each of these options in turn (of course, we are free to use combinations of the methods below):

\medskip\noindent
$\bullet$ {\it Discrete symmetries}: Imposing discrete symmetries can be a successful strategy in some theories (related comments have appeared in \SundrumGV). However, since the UV SCFT is supersymmetric and the soft terms manifestly break SUSY, this amounts to an assumption about the properties of the sector that mediates SUSY breaking. For example, consider SQCD. We can easily forbid UV soft terms as in \gencurrUVx\ that are proportional to the bottom component of the baryon current superfield by imposing the ${\bf Z_2}$ symmetry under which $Q\leftrightarrow\tilde Q$ (and the vector fields transform appropriately). In this case there is a simple mediation mechanism that respects this parity symmetry: gauge mediation using a weakly gauged diagonal subgroup of the $SU(N_f)_L\times SU(N_f)_R$ flavor symmetry. The price we pay for this is having to assume more about other sectors of particle physics.

\medskip\noindent
$\bullet$ {\it Gauging symmetries:} Turning on a gauge coupling for some of the global symmetries of the UV theory is useful if we would like to produce an emergent SUSY sector of the Standard Model (SM)---e.g., a sector with light stops---since we need to make contact with the SM gauge group. Turning on such couplings for a non-abelian group has the added advantage of forbidding (by gauge invariance) soft terms proportional to the bottom components of the corresponding current superfields (this fact has also been commented on in \SundrumGV). However, it leaves intact the possibility of soft term deformations proportional to the bottom components of abelian current superfields (these objects are gauge invariant under the abelian symmetry).

\medskip\noindent
$\bullet$ {\it Additional relevant deformations:} We can consider breaking some of the UV SCFT symmetries by turning on additional relevant (or dangerously irrelevant) superpotential deformations. This possibility often leads down a rabbit hole: if the deformations are small, we do not improve matters since we can simply expand in the deformation parameter. If the deformations are large, however, the soft terms corresponding to the currents that are broken by these deformations generally undergo incalculable RG evolution \AbelWV\ (this can also happen for strong additional gauging). 

\medskip\noindent
$\bullet$ {\it Engineered UV SCFT with minimal symmetry content:} Ideally we would find a UV SCFT with a global symmetry group, $G_{UV}$, that contains a minimal embedding of the SM gauge group, $G_{\rm SM}=SU(3)\times SU(2)\times U(1)_Y$, and an additional global $U(1)$ symmetry
\eqn\Grest{
G_{UV}\supseteq U(1)\times G_{SM}~.
}
The additional $U(1)$ factor is broken by the (marginally) relevant deformation that starts the RG flow, and the corresponding current mixes with the $U_{\rm vis}$ operator. One way to obtain a particularly restrictive UV setup is to find a theory with $G_{UV}=U(1)\times SU(5)$ and turn on a gauge coupling for the $SU(5)$. In this case, the universal soft term (of UV dimension two) is the canonical one given by 
\eqn\defUVii{
\delta S_{UV}|_{\rm SSB}=-\int d^4x\lambda\cdot U_{\rm vis}^{UV}~.
}
Such a setup may even be engineered in an asymptotically free gauge theory with a single representation of sufficient multiplicity (see \globalsymmpre; the missing $U(1)$ factor is the broken---in this case by anomaly---$U(1)$ factor we have just discussed). We could then perhaps imagine the states of phenomenological interest appearing as magnetic degrees of freedom in the IR of the resulting theory with suppressed soft masses. This scenario is potentially interesting since it requires us to make contact with GUT physics and duality (see \AbelBJ\ for an exploration of this subject). Of course, we can apply these lessons to other types of constructions as well: it may be possible to embed $U(1)_Y$ in a (weakly gauged) non-abelian symmetry without assuming a GUT structure in such a way that \defUVii\ is still the universal soft deformation (of dimension two) allowed in the UV. We should also remark that it may not be possible to avoid introducing additional global symmetries in a working model. In this case, we can try to use discrete symmetries, as discussed above, in order to forbid some of the corresponding soft terms. Note, however, that these global symmetries may often be broken by the additional relevant deformations that generate the IR states we want for phenomenology (e.g. light stops) and eliminate exotics. As we have outlined above, breaking additional UV global symmetries (beyond the $U(1)$ that mixes with $U_{\rm vis}$) may lead to incalculable RG evolution for the corresponding soft terms. In order to help avoid this situation, it would be of interest to find a theory in which we can encode the particle physics into the exactly marginal deformations of the UV SCFT (more precisely, into the subset of such deformations that do not break any of the global symmetries of the UV SCFT).\foot{Exactly marginal deformations move us along the conformal manifold. In the language of \GreenDA, these deformations consist of the marginal deformations modulo the action of the complexified global symmetry group, $G_{UV}^{\bf C}$ (this result can be appropriately generalized if there are also free gauge fields in the theory \GreenDA). A natural (although sometimes trivial) subset of these deformations does not break any of the UV SCFT symmetries. We would like to encode the particle physics into the sub-manifold spanned by these deformations.} In this case, it may be possible to engineer phenomenologically viable theories in which only the $U(1)\subset G_{UV}$ factor in \Grest\ would be broken away from the UV fixed point. This might lead to a calculable (since we avoid troublesome additional broken currents) and UV-insensitive RG flow with light stops in the IR.

\medskip

\subsec{Consequences for models of light sparticles: IR aspects}
Thus far we have emphasized aspects of the UV problem of engineering models of light sparticles. In the IR there are additional subtleties. One important issue we have briefly discussed in previous sections is the fact that $U^{IR}_{\rm vis}=0$ does not mean that SUSY breaking decouples in the IR. Indeed, if the flow of $U_{\rm vis}\to0$ is logarithmic, the suppression of IR masses is only logarithmic. This happens whenever there are accidental bosonic symmetries in the IR that mix with $U_{\rm vis}$ away from the IR fixed point. In free theories like SQCD in the free magnetic range ($N_c+1<N_f\le 3N_c/2$) we typically have such mixing since the approach to the free theory is controlled by marginally irrelevant operators (note that away from $N_f=3N_c/2$ the situation is worse because there is non-zero mixing of $U_{\rm vis}$ with accidental symmetries {\it at} the IR fixed point). Bearing this in mind, we consider two possible IR scenarios

\medskip\noindent
$\bullet$ {\it IR free theories:} Such theories have the advantage of being calculable. However, they typically come with marginally irrelevant interactions (some of which naturally play the role of Yukawa couplings) and so the IR SUSY breaking proportional to $U_{\rm vis}$ is typically suppressed at most only logarithmically (one should check numerical factors to ensure that this suppression suffices). In order to have more robust power-law suppression we need theories that behave like $\CL_{IR}=\int d^4\theta\left(\bar{\hat\Phi_a}\hat\Phi^a+\kappa\left(\bar{\hat{\Phi}_a}\hat\Phi^a\right)^2\right)$ in the deep IR. There should be some underlying symmetry principle at work that ensures this behavior. However, as we have described above, if this symmetry is a symmetry of the full theory (beyond just the two derivative theory), we are led to expect large SUSY-breaking masses corresponding to those symmetries.

\medskip\noindent
$\bullet$ {\it IR interacting theories:} If such theories have no accidental symmetries, they may provide robust power-law suppression of soft terms proportional to $U_{\rm vis}$. However, these theories tend to be less calculable. In addition, we should include a relevant deformation that eventually takes us out of the conformal phase \NelsonMQ. It may also be interesting to incorporate the operator dimension bounds of \refs{\RattazziPE\PolandWG\RattazziYC\VichiUX\PolandEY-\GreenNQ} in such an analysis.

\medskip
The final point is that it is imperative to develop technology to compute the next-order corrections in the small SUSY-breaking parameter. This need is especially acute in theories of emergent SUSY at leading order, since we may find unwanted tachyons (see, however, \AbelWV, for a use of tachyons in SQCD with a weakly gauged baryon number).

\newsec{Conclusion and discussion}
In this brief note we have highlighted how simple first principles of QFT---including unitarity and 't Hooft anomaly matching---(and a conjectured principle about the RG flow) can constrain theories of emergent SUSY. We hope that these constraints lead to robust model building for theories of light stops and Higgses. In particular, it would be of great interest to see whether one can construct a theory that is insensitive to UV physics (e.g., the mediating sector), robust in suppressing IR masses, and also calculable. In order to find such a theory, it may be crucial to find the physical principles that allow us to follow more general SUSY breaking terms non-perturbatively.

\bigskip
\bigskip
\centerline {\bf Acknowledgments}
We are grateful to S. Abel, I. Antoniadis, and Z. Komargodski for previous collaboration on closely related issues as well as many stimulating discussions directly related to the content of the present paper. We would also like to thank G. F. Giudice and N. Seiberg for interesting comments and discussions. This work was supported in part by the European Commission under the ERC Advanced Grant 226371 and the contract PITN-GA-2009-237920.

\vfill\eject
\appendix{A}{Proof of corollary for asymptotically free theories}
Let us now complete the proof of the corollary mentioned in the text. We reproduce the main statement below for ease of reference:

\medskip\noindent
{\bf Corollary}: Consider the set of asymptotically free theories with simple gauge group and vanishing superpotential such that \rvisIRi\ holds. If such a theory has a set of non-anomalous global symmetries, $\hat J_a$, then it follows that a non-trivial subgroup of this symmetry will not decouple in the IR.

\medskip\noindent
{\bf Proof:} First note that any UV theory of this type can be described as follows: it has a gauge group, $G$, and representations, $r_i$, of multiplicities, $n_i$ ($i=1,..., s$). The dimensions of the representations are $d_i>0$, and the Dynkin indices are $T_i> 0$. The global symmetry group is
\eqn\globalsymm{
S=U(1)_R\times U(1)^{s-1}\times\prod_{i=1}^sSU(n_i)~.
}
In this case, $U(1)_R$ anomaly freedom implies that
\eqn\UiRanomfree{
\sum_{i=1}^sn_i\left(R(r_i)-1\right)T_i+T_{\rm adj}=0~.
}
Note that $T_{\rm adj}>0$ since $G$ must be non-abelian. 

Let us now suppose that the unbroken currents have vanishing 't Hooft anomalies as in \tHooft. We will then find a contradiction with \UiRanomfree. To arrive at this contradiction, first consider the case of a single type of representation (i.e., $s=1$). In this case, if the representation of the gauge group has multiplicity one (i.e., $n_1=1$), there is no global symmetry. Therefore, consider the case $n_1>1$. Let $\hat J=SU(n_1)$ and impose the vanishing of the $\Tr R\hat J\hat J$ anomaly, i.e.
\eqn\TrRJJzero{
n_1(q_{R}(r_1)-1)\cdot{1\over2}\cdot d_1=0~.
}
It follows that $q_{R}(r_1)=1$. Plugging this result back into \UiRanomfree\ (with $s=1$), we find a contradiction since $T_{\rm adj}>0$. Therefore, it cannot be  that $\hat J\to0$ in the deep IR.

Next, consider the case that $s>1$ with $n_i=1$ for all $i=1, ..., s$ and let $\hat J_a=U(1)_a$ for all $a=1,..., s-1$. Let us impose that the $\Tr R^2\hat J_a$ and $\Tr \hat J_b^2\hat J_a$ anomalies vanish
\eqn\trui{
\sum_{i=1}^s(q_R(r_i)-1)^2\cdot d_i\cdot q_{i, a}=0, \ \ \ \sum_{i=1}^sq_{i,b}^2\cdot d_i\cdot q_{i, a}=0~,
}
where $q_{i,a}$ is the charge of representation $r_i$ with respect to $\hat J_a$. The $q_{i,a}$'s form a basis for an $s-1$ dimensional subspace of ${\bf R^s}$. Therefore, the vectors $(q_R(r_i)-1)^2\cdot d_i$ and $q^2_{i,b}\cdot d_i$ must sit in the one dimensional space orthogonal to the $q_{i,a}$'s. Vanishing of the $\Tr \hat J_a$ anomalies imposes
\eqn\Jaanom{
\sum_{i=1}^sd_i\cdot q_{i,a}=0~,
}
and so it follows that $(q_R(r_i)-1)^2=c^2$ and $q^2_{i,b}=c_b^2$ for some constants, $c^2$ and $c_b^2$, that are independent of the representation, $r_i$. The gauge anomaly freedom of the $U(1)_a$ symmetries implies
\eqn\anomfree{
\sum_{i=1}^sT_i\cdot q_{i,a}=0~.
}
Therefore, we can conclude that $T_i=b \cdot d_i$ for some constant $b>0$ that is independent of the representation.

Now, consider imposing $\Tr RU(1)_a^2$ anomaly freedom
\eqn\JUUi{
\sum_{i=1}^s(q_R(r_i)-1)\cdot d_i\cdot q^2_{i,a}=0~.
}
From this equation, it follows that
\eqn\JUUicons{
\sum_{i=1}^s(q_R(r_i)-1)\cdot d_i=0~,
}
and using the fact that $T_i=b\cdot d_i$ we then find a contradiction with \UiRanomfree\ since $T_{\rm adj}>0$.

Finally, consider the more generic case with $s>1$ and at least one $n_i>1$. Imposing the vanishing of the non-abelian anomalies $\Tr R SU(n_i)SU(n_i)$ results in
\eqn\TrRJJzero{
n_i(q_R(r_i)-1)\cdot{1\over2}\cdot d_i=0~,
}
for any $r_i$ with $n_i>1$. In particular, we see that such representations necessarily have $R$ charge $+1$ and therefore do not contribute to \UiRanomfree. Therefore, we are left to compute the contributions from the representations having $n_i=1$. To that end, let us impose the $s-1$ equations coming from the vanishing of the $\Tr RU(1)_a^2$ anomalies
\eqn\trui{
\sum_{i=1}^{s}(q_R(r_i)-1)^2\cdot n_i\cdot d_i\cdot q_{i, a}=0~,
}
As before, the $q_{i,a}$ form a basis for an $s-1$ dimensional subspace of ${\bf R^s}$. Therefore, the vector $(q_R(r_i)-1)^2\cdot n_i\cdot d_i$ must sit in the one dimensional space orthogonal to the $q_{i,a}$'s. Further imposing the vanishing of the $\Tr U(1)_a$ anomalies 
\eqn\trui{
\sum_{i=1}^sn_i\cdot d_i\cdot q_{i,a}=0~,
}
tells us that $(q_R(r_i)-1)^2=c^2$ for some constant $c^2$. However, from the vanishing of the non-abelian anomalies we know that $c^2=0$. Therefore, we again find a contradiction with \UiRanomfree, and the currents cannot flow to zero. {\bf q.e.d}

\listrefs
\end